\documentclass[prl,a4paper,twocolumn,amsmath,superscriptaddress]{revtex4-2}
\usepackage{graphicx}
\usepackage{epstopdf}
\usepackage{amsmath}
\usepackage{multirow}
\usepackage{color}

\begin{document}

\title{Polaritonic states of matter in rotating cavity}

\author{Lorenz S. Cederbaum}
\email[]{E-mail:  Lorenz.Cederbaum@pci.uni-heidelberg.de}
\affiliation{Theoretische Chemie, Physikalisch-Chemisches Institut, Universit\"at Heidelberg, Im Neuenheimer Feld 229, Heidelberg D-69120, Germany}
\date{\today}

\begin{abstract}
 
The interaction of quantum light with matter like that inside a cavity is known to give rise to mixed light-matter states called polaritons. We discuss the impact of rotation of the cavity on the polaritons. It is shown that the number of polaritons increases due to this rotation. The structure of the original polaritons is modified and new ones are induced by the rotation that strongly depend on the angular velocity and the choice of axis of rotation. In molecules the rotation can change the number of light-induced conical intersections and their dimensionality and hence strongly impact their quantum dynamics. General consequences are discussed.

\end{abstract}

\maketitle

The coupling of matter excitations with quantized radiation field like that inside a cavity has become an extensive field of research. Most importantly, this coupling leads to the formation of hybrid light-matter states called polaritons or polaritonic states which open up many possibilities to enhance or suppress available mechanisms and to mediate new ones, and to manipulate and control the properties of the matter. The list of work published in this field has grown up immensely. Here, we just mention exemplary the possibilities to control chemical reactions by varying the properties of the quantized field \cite{Cavity_Chem_Reac_3,Cavity_Chem_Reac_4,Cavity_Chem_Reac_5,Cavity_Chem_Reac_6}, to control photochemical reactivity \cite{Cavity_Chem_Reac_1,Cavity_Chem_Reac_2},   enhance energy transfer \cite{Cavity_CT_3,Cavity_ET_1} and charge transfer \cite{Cavity_CT_1,Cavity_CT_2,Cavity_CT_3,Cavity_CT_4} as well as of molecular non-adiabatic processes \cite{Cavity_Chem_Reac_5,Cavity_Non_Adiab_1,Cavity_Non_Adiab_2,Cavity_LICI_1,Cavity_LICI_3,Cavity_LICI_4,Cavity_Coll_CI}.  

We study the impact of rotation of a cavity on the polaritonic states of atoms and molecules interacting resonantly with the cavity and show below that the effects induced by the rotation can be rather severe. Since it is the first work on the subject, we concentrate here on the fundamental aspect of this impact and do not discuss technical details concerning the choice of the cavity and the possible back-action of the rotation on the cavity. The physics of rotating systems has attracted much attention in many contexts. For instance, ultracold bosonic gases under rotation probe various phenomena of correlated quantum systems including quantized vortices \cite{BEC_Rotation_Vortices_1,BEC_Rotation_Vortices_2,BEC_Rotation_Vortices_3}, the enhancement of many-body effects in the excitation spectrum \cite{BEC_Rotation_Beinke}, and much more. Extensive literature on rotating Bose-Einstein condensates can be found in reviews \cite{BEC_Rotation_Review_1,BEC_Rotation_Review_2,BEC_Rotation_Review_3}. Other subjects of interest in rotating systems comprise geometric quantum
phases \cite{Rotation_Geometric_Phase_1,Rotation_Geometric_Phase_2}, quantum rings \cite{Rotation_Quantum_Rings_1,Rotation_Quantum_Rings_2}, electroweak interactions \cite{Rotation_Electroweak}, spintronics \cite{Rotation_Spintronics_1,Rotation_Spintronics_2} and induced electric dipole moment interacting with external fields \cite{Rotation_Induced_Dipole}. 

We consider an ensemble of $N$ non-interacting identical atoms or molecules in a cavity with a quantized light mode (cavity mode) of frequency $\omega_c$ and polarization direction $\vec{\epsilon}_c$ and let the cavity rotate with a uniform angular velocity $\Omega$ around an axis $u$. The total Hamiltonian $H_L$ of the ensemble plus rotating cavity and the solution of the time-dependent Schr\"odinger equation $\Psi_L$ in the laboratory frame can be expressed by the respective quantities in the rotating frame ($RF$) by employing the rotation operator $\hat{O}_u = \exp(-i\Omega\vec{u}\cdot\vec{\mathcal{L}}t/\hbar)$, where $\vec{u}$ is the unit vector along the rotation axis and $\vec{\mathcal{L}}$ is the total angular momentum of all particles. Inserting $H_{L} = \hat{O}_u H_{RF} {\hat{O}_u}^\dagger$ and $\Psi_{L} = \hat{O}_u \Psi_{RF}$ into the time-dependent Schr\"odinger equation in the laboratory frame, one readily finds 

\begin{align}\label{SE_RF}
i\hbar	\frac{\partial\Psi_{RF}}{\partial t} =  [H_{RF} - \vec{\Omega}\cdot\vec{\mathcal{L}}] \Psi_{RF}
\end{align}	
for the time-dependent Schr\"odinger equation in the rotating frame, where $\vec{\Omega} = \Omega\vec{u}$. In this frame the Hamiltonian $H_{RF}$ is time-independent and just the total Hamiltonian of the ensemble-cavity system with a non-rotating cavity and reads \cite{Cohen_Tannoudji_Book,Faisal_Book,Feist_PhysRevX,Oriol_Cavity_CP}:

\begin{align}\label{Ensemble-Cavity-Hamiltonian}
	H_{RF} = H_e + \hbar\omega_c\hat{a}^\dagger \hat{a} + g_0 \vec{\epsilon}_c\cdot \vec{d}(\hat{a}^\dagger + \hat{a}),
\end{align}
where $H_e = \sum_{i=1}^{N} H_i$ is the Hamiltonian of the ensemble, $\vec{d} = \sum_{i=1}^{N} \vec{d}_i$ is the total dipole operator of the ensemble and $g_0$ is the coupling strength between the cavity and the atoms. The quadratic dipole self-energy term is neglected as it is only of relevance for very strong coupling.  

Let us start the discussion with a single atom and assume its excitation energy to be resonant with the cavity mode. For simplicity of presentation, we choose the angular momentum of the atom in its ground state $\phi_{l,m}$ to be $l=0$ and $l=1$ in its excited state $\psi_{l,m}$ and take the ground state energy to be the zero of the energy scale. As usual, we resort to the single-excitation space where the contributing states are the atom in its ground state and the cavity in a single photon state denoted by $\phi_{0,0}1_c$ and the atom in its excited state with the cavity having zero photons denoted analogously. Without rotation, the Hamiltonian of the atom-cavity system is known to have two polaritonc solutions $\Phi_{up/lp} = (\phi_{0,0}1_c \pm \psi_{1,0}0_c)/\sqrt{2}$ with energies $E_{up/lp} = \hbar\omega_c \pm g$ called upper and lower polaritons , where $g=g_0d_z$ and $d_z$ is the $Z$ component of the transition dipole. Note that owing to the dipole interaction, only the $m=0$ state of the excited states contributes. 

The situation changes once the cavity rotates. The angular momentum operator of the center of mass of the atom contributes the same to all states and can be ignored and one is left in Eq. (\ref{SE_RF}) with the electronic angular momentum operator $\vec{L}$. One notes that, in general, all three excited states $\psi_{1,m}$, $m=0,\pm 1$ of the atom must be considered now. Only if the rotation axis $u$ coincides with the polarization direction of the cavity, the $m= \pm 1$ states decouple and can be ignored. Next we call the polarization direction $Z$ and rotate the cavity in the $XY$-plane. In the space spanned by $\phi_{0,0}1_c$ and the three $\psi_{1,m}0_c$ states, the Hamiltonian $\mathcal{H}  =  [H_{RF} - \vec{\Omega}\cdot\vec{L}]$ is represented by the matrix

\begin{align}\label{XY-Hamiltonian}
\mathbf{\mathcal{H}}
=
\begin{pmatrix}
	\hbar\omega_c & g & 0 & 0\\
	g & \hbar\omega_c & -\hbar\Omega_- & -\hbar\Omega_+\\
	0 & -\hbar\Omega_+ & \hbar\omega_c & 0\\
	0 & -\hbar\Omega_- &  0 & \hbar\omega_c
\end{pmatrix}
,
\end{align}
where the ladder operators $L_{\pm} = (L_x \pm iL_y)/\sqrt{2}$ and $\Omega_xL_x + \Omega_yL_y = \Omega_+L_+ + \Omega_-L_-$ have been introduced.  The above matrix is hermitian as $\Omega_-^\dagger = \Omega_+$. We see that the cavity mode couples directly only to the $m=0$ excited state, but this state couples to the $m=\pm1$ states.	

To proceed, we diagonalize the lower right $3\times3$ block of the above matrix and the resulting eigenvectors can be used as a basis for the excited electronic manifold in which the Hamiltonian $\mathcal{H}$ becomes an arrowhead matrix \cite{arrowhead_1,arrowhead_2,arrowhead_3} which is particularly helpful when extending the problem to $N$ atoms. The new basis can be given explicitly. The state $\psi_0 = (\Omega_+\psi_{1,1}0_c - \Omega_-\psi_{1,-1}0_c)/\Omega$ has the energy $\hbar\omega_c$ and decouples from all the other states including the cavity one photon state $\phi_{0,0}1_c$ and hence can be addressed as a dark state. It should be noticed that the dark state is an entangled state due to rotation of the cavity. The remaining states read $\psi_{\pm} = [\psi_{1,0}0_c \mp  (\Omega_+\psi_{1,1}0_c + \Omega_-\psi_{1,-1}0_c)/\Omega]/\sqrt{2}$. Without the dark state, the matrix Hamiltonian in the $\phi_{0,0}1_c$ and $\psi_{\pm}$ basis now takes on the simple appearance
\begin{align}\label{XY-Hamiltonian_2}
	\bf{\mathcal{H}}
	=
	\begin{pmatrix}
		\hbar\omega_c & g/\sqrt{2} & g/\sqrt{2}\\
		g/\sqrt{2} & \hbar(\omega_c+\Omega)  & 0\\
	    g/\sqrt{2} & 0 & \hbar(\omega_c-\Omega)\\
		\end{pmatrix}
	,
\end{align}
where we remind that $\Omega$ is the angular velocity of the cavity. The eigenvalues and eigenvectors are easy to obtain. Interestingly, one of the eigenvalues is again just $\hbar\omega_c$, but in contrast to the above mentioned state, this state in not a dark state. Apart from a normalization constant, it reads $[(\hbar\Omega\sqrt{2}/g)\phi_{0,0}1_c + \psi_{-} - \psi_{+}]$, i.e., this state is a new kind of a polariton and is solely induced by the rotation of the cavity. Although modified by the rotation, the other two states are reminiscent of the upper and lower polaritons present in the non-rotating cavity. Their energies are also modified by the rotation giving $E_{up/lp} = \hbar\omega_c \pm [\hbar^2\Omega^2 +g^2]^{1/2}$. 

What happens when more atoms are present in the cavity? For a non-rotating cavity and $N$ atoms it is well known that there are two polaritonic states with energy $E_{up/lp}(N) = \hbar\omega_c \pm \sqrt{N}g$ and $N-1$ dark states at $\hbar\omega_c$. For a rotating cavity there are $3N+1$ states playing a role as each atom contributes 3 excited states. As we have seen above, for each atom one dark state entangled by the rotation can be decoupled. To investigate the remaining $2N+1$ states, one can extend the matrix Hamiltonian in Eq. (\ref{XY-Hamiltonian_2}) by repeating $N$ times the lower right $2\times2$ block along the diagonal and extending the first row and column by $g/\sqrt{2}$ coupling elements. The resulting arrowhead matrix can be solved using methods described in \cite{arrowhead_1,arrowhead_2,arrowhead_3}. To find the polaritonic states, one can use the Dyson-like equation $E-\hbar\omega_c = g^2N/2[1/[E-\hbar(\omega_c+\Omega)] +  1/[E-\hbar(\omega_c-\Omega)]]$ which readily leads to the eigenenergies
\begin{align}\label{XY-Energies}
\nonumber & E_{up/lp} = \hbar\omega_c \pm [(\hbar\Omega)^2 + Ng^2]^{1/2},\\&
E_0 = \hbar\omega_c.
\end{align}
As we found above for a single atom, there are three polaritons, two of them reminiscent of those in a non-rotating cavity and one at $\hbar\omega_c$ induced by the rotation. In addition, there are $N-1$ dark states at energy $\hbar(\omega_c+\Omega)$ and $N-1$ dark states at energy $\hbar(\omega_c-\Omega)$. I.e., this kind of dark states are doubled and shifted by the rotational frequency compared to the case of a non-rotating cavity.

Above, we have seen that there is no impact of rotation for a cavity rotating around its polarization direction and that the impact is severe when the rotation axis is perpendicular to the polarization direction. What if the unit vector $\vec{u}$ along the rotation axis has components in all directions? Then, indeed, also $\Omega_z$ contributes and makes the outcome even more intricate. We first return to a single atom and to the Hamiltonian matrix in Eq. (\ref{XY-Hamiltonian}). The action of $-\Omega_zL_z$ has to be added to this matrix. I.e., $-\hbar\Omega_z$ has to be added to the third element along the diagonal and $\hbar\Omega_z$ to the last one. Now, one can proceed as done above and diagonalize the lower right $3\times3$ block of the Hamiltonian matrix and use the resulting eigenvectors as a new basis to represent the Hamiltonian matrix. The resulting Hamiltonian matrix takes on the following appearance 
\begin{align}\label{XYZ-Hamiltonian}
	\bf{\mathcal{H}}
	=
	\begin{pmatrix}
		\hbar\omega_c & \tilde{g}\Omega_{xy}/\Omega & \tilde{g}\Omega_{xy}/\Omega & g\Omega_{z}/\Omega\\
		\tilde{g}\Omega_{xy}/\Omega & \hbar(\omega_c+\Omega) & 0 & 0\\
		\tilde{g}\Omega_{xy}/\Omega & 0 & \hbar(\omega_c-\Omega) & 0\\
		g\Omega_{z}/\Omega & 0 &  0 & \hbar\omega_c
	\end{pmatrix}
	,
\end{align}
where $\tilde{g}=g/\sqrt{2}$ and $\Omega_{xy} = (\Omega_x^2 + \Omega_y^2)^{1/2}$ is the effective angular velocity in the $XY$-plane. Note that $\Omega^2 = \Omega_{xy}^2 + \Omega_{z}^2$. By putting $\Omega_{z}$ to zero, one of the states decouples and becomes a dark state of energy $\hbar\omega_c$ and one immediately recovers the matrix Hamiltonian in Eq. (\ref{XY-Hamiltonian_2}) for the polaritons in the cavity rotating around an axis in the $XY$-plane.  

A major outcome of the above discussion is that if the axis of rotation is neither parallel to the polarization axis nor in the plane perpendicular to it, one obtains four polaritonic states instead of two in the non-rotating cavity. Two of the polaritonic states are solely induced by the rotation.

If we have $N$ atoms in the rotating cavity, the Hamiltonian matrix can be obtained from that in Eq. (\ref{XYZ-Hamiltonian}) by repeating $N$ times the lower right $3\times3$ block along the diagonal and extending the first row and column by $N$ times by coupling trio $\tilde{g}\Omega_{xy}/\Omega, \tilde{g}\Omega_{xy}/\Omega, g\Omega_{z}/\Omega$. Again, the resulting arrowhead matrix can be solved and the polaritonic states follow from the Dyson-like equation $E-\hbar\omega_c = ({\tilde{g}\Omega_{xy}/\Omega})^2N[1/[E-\hbar(\omega_c+\Omega)] +  1/[E-\hbar(\omega_c-\Omega)]]+({g\Omega_{z}/\Omega})^2N/[E-\hbar\omega_c]$ which can be solved explicitly giving the four energies of the polaritons:
\begin{align}\label{XYZ-Energies}
\nonumber E_{polariton} = \hbar\omega_c \pm \frac{1}{\sqrt{2}}\big\{[(\hbar\Omega)^2 + Ng^2] \\ \pm \{[\hbar^2(\Omega_{xy}^2-\Omega_{z}^2) + Ng^2]^2 + 4\hbar^4\Omega_{xy}^2\Omega_{z}^2\}^{1/2}\big\}^{1/2}.
\end{align}
The different impact of $\Omega_{z}$ and $\Omega_{xy}$ is clearly seen in the above energies of the polaritons. The four polariton branches are accompanied by $3N-3$ dark states, $N-1$ at energy $\hbar\omega_c$ and $N-1$ at each of the energies $\hbar(\omega_c\pm\Omega)$. 

We now turn to molecules. Due to the additional vibrational and rotational degrees of freedom, the treatment of molecules is more involved. A particularly intriguing quantity in polyatomic molecules are conical intersections (CIs) which lead to singular coupling between the electronic and nuclear degrees of freedom \cite{Conical_Intersections_Review_1,Conical_Intersections_Book_1}.    
Classical laser fields can induce new CIs, named light-induced CIs (LICIs) even in a single diatomic molecule \cite{LICI_1,LICI_2,LICI_3}. A quantized radiation field like that in a cavity also induces LICIs in a diatomic with new implications on their dynamic properties \cite{Cavity_LICI_1,Cavity_LICI_2,Cavity_LICI_3} and, of course, also in polyatomics \cite{Cavity_LICI_3,Cavity_LICI_4}. New types of intersections appear when more molecules are subject to the same quantized field where the molecules interact with each other via the field, like the collective conical intersection which gives rise to unusual dynamics \cite{Cavity_Coll_CI}.

Each type of molecule has to be treated by itself, as its structure and the nature of its electronic ground and excited states play a role. To be specific we consider a closed-shell homonuclear diatomic molecule in a cavity where its ground state with one cavity photon $\phi_{\Sigma}1_c$ interacts with a doubly degenerate $\psi_{\Pi}0_c$ state of $\Pi$ symmetry with zero photons. This example has the ingredients needed for understanding how to treat other systems. We proceed in three steps. In the first step we freeze the nuclear motion of the molecule and consider the electronic-cavity matrix Hamiltonian for the system in a non-rotating cavity, i.e., the internuclear distance $r$ and the rotational angles $\theta$ and $\varphi$ are kept fixed and have the role of parameters. In the space of the above states this Hamiltonian is a $3\times3$ matrix, but depending on the polarization of the cavity, one can transform the components of the degenerate $\psi_{\Pi}0_c$ state such that one component decouples resulting effectively in a $2\times2$ matrix, see, e.g., \cite{LICI_MgO_Communication}. As we shall see below, in a rotating cavity the Hamiltonian cannot be reduced and stays a $3\times3$ matrix. 

Choosing $\psi_{\Pi_\pm} = (\psi_{\Pi_x} \pm \psi_{\Pi_y})/\sqrt{2}$, one obtains the following electronic-cavity matrix Hamiltonian for the system in a non-rotating cavity
\begin{align}\label{Sigma_Pi-Hamiltonian_no_Rotation}
  \begin{pmatrix}
  V_{\Sigma}(r) + \hbar\omega_c & g(r)\sin(\theta)/\sqrt{2} & g(r)\sin(\theta)/\sqrt{2}\\
  g(r)\sin(\theta)/\sqrt{2} & V_{\Pi}(r)  & 0\\
  g(r)\sin(\theta)/\sqrt{2} & 0 & V_{\Pi}(r)
  \end{pmatrix}
.
\end{align}  
$V_{\Sigma}$ and $V_{\Pi}$ are the potential energies corresponding to the $\phi_{\Sigma}$ and $\psi_{\Pi}$ electronic states, $g(r)=g_0d_{_{\Sigma\Pi}}(r)$, where $d_{_{\Sigma\Pi}}$ is the transition dipole moment between the two electronic states $\phi_{\Sigma}$ and $\psi_{\Pi_x}$ and depends on the interatomic distance, and $d_{_{\Sigma\Pi}}\sin(\theta)$, where $\theta$ is the angle between the molecular axis and the cavity polarization, is the projection of the moment on the polarization direction. It is easily seen that a rotation of the degenerate $\Pi$-states by $\pi/4$ leads to the decoupling of one of them from the $\phi_{\Sigma}1_c$ state.

In the second step we allow the cavity to rotate and consider the impact of adding $- \vec{\Omega}\cdot\vec{L}$ to the electronic-cavity Hamiltonian in Eq. (\ref{Sigma_Pi-Hamiltonian_no_Rotation}). We denote the resulting matrix Hamiltonian by $\mathbf{\mathcal{H}}_{ec}$ to distinguish it from that of the total Hamiltonian $\bf{\mathcal{H}}$ which includes the nuclear motion of the molecule in the rotating cavity. To bring a diatomic molecule into a general position in the cavity, we start with the molecule along the $Z$-axis and rotate it first by an angle $\theta$ around the $Y$-axis and then by an angle $\varphi$ around the $Z$-axis. The resulting rotation matrix $\bf{R}$ is
\begin{align}\label{Rotation_Matrix}
\mathbf{R}
=
\begin{pmatrix}
\cos(\varphi)\cos(\theta) & -\sin(\varphi) & -\cos(\varphi)\sin(\theta)\\
\sin(\varphi)\cos(\theta) & \cos(\varphi) & -\sin(\varphi)\sin(\theta)\\
\sin(\theta) & 0 & \cos(\theta)
\end{pmatrix}
.
\end{align} 
The states $\psi_{\Pi_\pm}$ are transformed by the rotation to become $U(\mathbf{R})\psi_{\Pi_\pm}(\vec{r}_e) = \psi_{\Pi_\pm}({\bf{R}}^{-1}\vec{r_e})$, where $\vec{r_e}$ indicates the electronic coordinates. To determine the matrix elements of the electronic angular momentum between the rotated states, we use the fact that the angular momentum is a vector and hence \cite{QM_II}:    
\begin{align}\label{Rotation of L}
\langle\psi_{\Pi_\pm}|U^\dagger(\mathbf{R})\vec{L}U(\mathbf{R})|\psi_{\Pi_\pm}\rangle = \langle\psi_{\Pi_\pm}|\mathbf{R}\vec{L}|\psi_{\Pi_\pm}\rangle .
\end{align}
This implies that the matrix element of each component of the angular momentum vector between any two rotated states can be expressed as a linear combination of the matrix elements of all the components between the respective unrotated states. The latter are usually known. 

In the present case of a diatomic molecule only the diagonal matrix elements of $L_z$ do not vanish and as the $\psi_{\Pi_\pm}$ are eigenstates of $L_z$, $L_z\psi_{\Pi_\pm} = \pm\psi_{\Pi_\pm}$, the resulting electronic-cavity Hamiltonian $\mathbf{\mathcal{H}}_{ec}$ takes on the following appearance
\begin{widetext}
\begin{align}\label{Sigma_Pi-Hamiltonian_Rotation}
\mathbf{\mathcal{H}}_{ec}
=
\begin{pmatrix}
V_{\Sigma}(r) + \hbar\omega_c & g(r)\sin(\theta)/\sqrt{2} & g(r)\sin(\theta)/\sqrt{2}\\
g(r)\sin(\theta)/\sqrt{2} & V_{\Pi}(r) - f(\varphi)\sin(\theta) + \cos(\theta)\hbar\Omega_z  & 0\\
g(r)\sin(\theta)/\sqrt{2} & 0 & V_{\Pi}(r) + f(\varphi)\sin(\theta)  - \cos(\theta)\hbar\Omega_z
\end{pmatrix}
,
\end{align}
\end{widetext}
where $f(\varphi) = \hbar(\Omega_+ e^{i\varphi} + \Omega_- e^{-i\varphi})\sqrt{2}$ is a real function of $\varphi$. Two consequences are immediately seen: Even if the rotation is only around the polarization axis, three polaritons emerge, and, for a general rotation, the azimuthal angle enters and the polaritons now depend on the three parameters $r$, $\theta$ and $\varphi$.

As can be seen from Eq. (\ref{Sigma_Pi-Hamiltonian_no_Rotation}), two LICIs between the potential energy surfaces (eigenvalues) of the Hamiltonian appear in a non-rotating cavity for $\theta=0,\pi$ at the value of $r$ where the condition $V_\Sigma(r) + \hbar\omega_c= V_\Pi(r)$ is met, see also \cite{LICI_MgO_Communication}. If the cavity rotates around $Z$, the number of LICIs doubles. The condition $\theta=0,\pi$ still applies for all LICIs. One pair of LICIs appears at $r_+$ and one at $r_-$ where $V_\Sigma(r_\pm) + \hbar\omega_c= V_\Pi(r_\pm) \pm \hbar\Omega_z$ are fulfilled, respectively. If the cavity rotates around a general axis, the potential surfaces become three dimensional as they are found to also depend on the angle $\varphi$. In three dimensions, conical intersections are not anymore distinct points but one-dimensional hypersurfaces in nuclear coordinate space \cite{Conical_Intersections_Review_1,Conical_Intersections_Book_1}. Indeed, for a rotating cavity around a general axis, the conditions for a LICI are the same as above and each of the four LICIs exists on a seam $2\pi \geq \varphi \geq 0$. As the presence of LICIs has substantial effect on the dynamics of molecules, the present result shows that one can investigate the impact of a seam of LICI for a diatomic molecule by rotating the cavity.
   
In the third step we take into account the vibrational and rotational motions and consider the total Hamiltonian $\mathcal{H} = H_{RF} - \vec{\Omega}\cdot\vec{\mathcal{L}}$ including the respective kinetic energies and, in addition, to the electronic angular momentum $\vec{L}$ also the common rotational angular momentum $\vec{L}_{\varphi\theta}$ of the molecule in $\vec{\mathcal{L}} = \vec{L} + \vec{L}_{\varphi\theta}$. The kinetic energy operator of a diatomic is $-\frac{\hbar^2}{2\mu}\frac{\partial ^2}{\partial r^2} + \frac{L_{\varphi\theta}^2}{2\mu r^2}$, where $\mu$ is the reduced mass of the molecule. Adding to it the impact $- \vec{\Omega}\cdot\vec{L}_{\varphi\theta}$ of the rotation of the cavity, gives rise to the final Hamiltonian of the molecule interacting with the rotating cavity 
\begin{align}\label{Final_Diatomic_Hamiltonian}
\mathbf{\mathcal{H}} = \left[-\frac{\hbar^2}{2\mu}\frac{\partial ^2}{\partial r^2} + \frac{(\vec{L}_{\varphi\theta}- \mu r^2 \vec{\Omega})^2}{2\mu r^2} - \frac{\mu r^2 \Omega^2}{2}\right] \mathbf{1} + \mathbf{\mathcal{H}}_{ec},
\end{align}
where $\mathbf{1}$ is a unity $3\times3$ matrix and $\mathcal{H}_{ec}$ is the electronic-cavity Hamiltonian in Eq. (\ref{Sigma_Pi-Hamiltonian_Rotation}). All the coordinates $r, \theta$ and $\varphi$ are now dynamic variables. The rotation of the cavity not only influences severely the structure of the polaritonic states as discussed above, the rotation and vibration of the molecules are affected as well.   

The rotation of a cavity changes the number of the polaritons found without rotation. Their number increases due to this rotation. The rotation modifies the structure of the original polaritonic states and induces new ones which strongly depend on the angular velocity and the choice of axis of rotation with respect to the polarization of the cavity mode. In molecules the rotation can change the number of light-induced conical intersections and their dimensionality and hence strongly impact the molecular dynamics. If the axis of rotation differs from the cavity polarization direction, a seam of LICIs appears even in a diatomic molecule. By changing the axis of rotation, this would allow the investigation of the quantum dynamics due to a LICI seam. 

The impact of rotation on the polaritons of an ensemble of $N$ non-interacting identical atoms has been investigated. The number of polaritonic branches is the same as for an individual atom, but the number of dark states increases compared to a non-rotating cavity by a factor 2 or 3 depending on the axis of rotation, and, interestingly, their energy is shifted by the rotation. The Hamiltonian matrix for a diatomic molecule can be straightforwardly extended to accommodate $N$ non-interacting identical ones similarly as done for the atoms. However, as discussed before \cite{Cavity_Coll_CI}, the matrix is an operator in the $N$-dimensional (in the present case $3N$-dimensional) nuclear coordinate space as each molecule possesses its own dynamical variables, and this makes the analysis rather intricate and is left for the future.      

The study can also be extended to polyatomic molecules. Here, one has to resort to three angles instead of the two for diatomics. In general, Euler angles are a good choice and with them a similar strategy as applied here for the diatomic molecules can be employed  \cite{QM_II,LICI_Polyatomic}. Additional extensions of interest include the investigation of other atomic electronic angular momenta, of atoms and molecules with spin-orbit coupling, and as a long-term project, of an ensemble of interacting atoms or molecules.


\begin{acknowledgements}
	The author thanks O. E. Alon and A. I. Kuleff for valuable contributions and A. Vib\'ok for convincing him that cavities are interesting. Financial support by the European Research Council (ERC) (Advanced Investigator Grant No. 692657) is gratefully acknowledged
\end{acknowledgements}

\bibliographystyle{apsrev}
\bibliography{biblio}


\end{document}